# Direct Observation of Mono-, Bi-, and Tri-layer Charge Density Waves in 1$T$-TaS$_2$ by Transmission Electron Microscopy without a Substrate

Daiki Sakabe[1,2], Zheng Liu[3,4], Kazumoto Suenaga[2,4], Keiji Nakatsugawa[1,2]

& Satoshi Tanda [1,2]

[1]Department of Applied Physics, Hokkaido University, Sapporo, 060-8628, Japan

[2]Center of Education & Research for Topological Science & Technology, Hokkaido University,

  Sapporo, 060-8628, Japan

[3]Inorganic Functional Materials Research Institute, AIST, Nagoya, 463-8560, Japan

[4]Nanomaterials Research Institute, AIST, Tsukuba, 305-8565, Japan

Correspondence: Satoshi Tanda (tanda@eng.hokudai.ac.jp)

  **Abstract**

Charge-density-waves (CDW) [1, 2] which occur mainly in low-dimensional systems have a macroscopic wave function similar to superfluids and superconductors. Kosterlitz-Thouless (KT) transition [3, 4] is observed in superfluids and superconductors, but the presence of KT transition in ultra-thin CDW systems has been an open problem.

We report the direct real-space observation of CDWs with new order states in mono-, bi-, and tri-layer 1T-TaS$_2$ crystal by using a low voltage scanning-transmission-electron-microscope (STEM) without a substrate. This method is ideal to observe local atomic structures and possible defects. We clearly observed that the mono-layer crystal has a new triclinic stripe CDW order without the triple **q** condition $\mathbf{q}_1 + \mathbf{q}_2 + \mathbf{q}_3 = 0$. A strong electron-phonon interaction gives rise to new crevasse (line) type defects instead of disclination (point) type defects due to the KT transition. These results reaffirm the importance of the electron-phonon interaction in mono-layer nanophysics.

1. Introduction

Dimensionality and topology are the most important parameters characterizing physical systems. For example, the integral and fractional quantum Hall effects (QHE) [5, 6] are observed only in two-dimensional systems such as metal-oxide-semiconductor field-effect transistors, GaAs/AlGaAs interfaces and graphene.[7] Moreover, the conductivity of the QHE is characterized by topological numbers. Dimensionality and topology are also of importance in the Kosterlitz-Thouless (KT) transitions.[3] The KT transition is a phase transition that occurs only in a two-dimensional XY model, and vortices or vortex-pairs (topological defects) are known to play an important role. KT transitions are observed in systems that have Macroscopic Wave Functions (MWF) such as two- dimensional superconductors [8] and ultrathin film superfluids.

Charge density wave (CDW) [1, 2] is a phenomenon in which dimensionality and topology are of particular importance. A CDW is a modulation of electric charge in low-dimensional conductors caused by electron-phonon coupling and has a MWF [9] such as superconductors, superfluids and quantum Hall liquids. Additionally, the properties of a CDW are changed depending on its topology [10–12].

Tantalum disulfide with hexagonally packed $TaS_6$ (1T-$TaS_2$ (Fig. 1(a)) is a typical CDW material. 1T-$TaS_2$ is a layered compound called a transition metal dichalcogenides ($MX_2$), which induces two-dimensional CDW with wave-number vectors $\mathbf{q}_i$ (i = 1, 2, 3) termed triple q (Fig.1(b)). $MX_2$ belongs to the group of van der Waals materials characterized by their layered crystalline structures. A CDW in this material takes four different states depending on temperature (Fig. 1(c)).

How do the CDW properties change when we use thin samples? There is a possibility that the properties differ from those in higher layers (bulks[13, 14,15]). There has been a recent debate on whether the commensurate CDW phase exists in ultrathin films of 1T-TaS$_2$[16-20]. However, the samples considered in these studies either contain a substrate or interacts with the surrounding environment, which inevitably alter the properties of the material. However, it is possible that different CDW phases emerge in standing-free ultrathin 1T-TaS$_2$ as the thickness decreases down to a single layer. For example, the KT transition may occur in CDW systems.

In this paper, we report the properties of 1T-TaS$_2$ ultra-thin films including a mono-layer. We obtained images of 1T-TaS$_2$ using scanning transmission electron microscopy (STEM). Fig. 2 clearly shows an image of the crystal structure. STEM is a measurement method that does not require a substrate and causes little or no damage to the samples. Therefore, STEM is well suited for the measurement of ultrathin samples. We were able to make mono-, bi-, and tri-layer 1T-TaS$_2$ samples using the exfoliation method (c.f. *Method* section) for the first time. Therefore we used this technique to observe formation of different CDW phases in tri-layer, bi-layer, and mono-layer samples.

## 2. Results
### 2.1 Tri-layer CDW

Figure 3(a) shows an STEM image of 1T-TaS$_2$ at room temperature. Figure 3(b) is a three-dimensional intensity plot of the area shown by the yellow frame in Fig. 3(a). As seen from the figure, the brown circles form hexagonal lattices. It is a feature of Commensurate (C-) CDW that super-lattices of C-CDW form such large hexagonal lattices (Fig. 3(d), Fig. 3(e)).

Figure 3(c) shows a Fourier transformed image (FTI) of Fig. 3(a), where certain satellite peaks are present. The appearance of powerful satellites is a feature of C-CDW state. The **q** vectors are calculated from Fig. 3(c) with the values of $/\mathbf{q}/ = (0.280(8) \pm 0.003)/\mathbf{a}^*/$ that is rotated $\psi = 13.3(7) \pm 1.2°$ with respect to the Bragg spots. The C-CDW satellites were confirmed in at least 10 other STEM images of 1T-TaS$_2$ samples (Supplementary material 1). These results are in good agreement with the reported values for the C-CDW state: $/\mathbf{q}/ = (0.277 \pm 0.003)/\mathbf{a}^*/$, $\psi = 13.90°$ [15]. From the above evidence, we surprisingly discovered that C-CDW occurs in tri-layer 1T-TaS$_2$ at room temperature. Fig. 3(f) shows the satellite patterns of the tri-layer sample. The triple $\mathbf{q}$ condition is clearly preserved.

## 2.2 Bi-layer CDW

Figure 4(a) shows a STEM image of bi-layer 1T-TaS$_2$ at room temperature. The CDW super-lattices in the bi-layer are different from those in the tri-layer. Figure 4(b) shows a FTI of Fig. 4(a). The CDW satellites are largely diffused. Thus, a super-lattice is also present in the bi-layer (Fig. 4(c), Fig. 4(d)) but it is not as clearly defined as in the case of the tri-layer. However, it is obvious that there are triple $\mathbf{q}$ vectors (Table. 1(b)). This CDW has anisotropic triple $\mathbf{q}$ vectors such as T-phase [14,15,21] and stripe phase [22]. Being these types, the CDWs form domains with domain walls corresponding to discommensuration [23]. Consequently, the system minimizes its energy by making super-lattices in these domains similar to those in the C-phase. For this reason, it is valid to consider that domains are formed in this sample. Figure 4(c) shows an enlarged view of the yellow frame in Fig. 4(a) with inverted contrast. The size of Fig. 4(c) is 100Å squared. The size is similar to the C-CDW domain size in NC or T-phase [14, 15, 24]. However, C-type super-lattices such as those in the tri-layer crystal are not observed in Fig. 4(c). Figure 4(d) shows the FTI of Fig. 4(c) and the result is similar to Fig. 4(b). We performed Fourier transforms (FT) at different locations in Fig. 4(a) and confirmed similar satellite patterns. The sizes of domains that exist in the bi-layer crystal appear to be smaller than those in the bulk material. Unlike the tri-layer sample, clear super-lattices and honeycomb lattices are not seen. In view of the **q** vector's length, the CDW state in bi-layer is similar to the stripe phase in 2H-TaSe$_2$ [22.] Fig. 4(f) shows the satellite patterns of the bi-layer sample. It is shown that the triple **q** conditions starts to be broken.

**2.3 Mono-layer CDW**

Figure 5(a) shows an image of mono-layer 1T-TaS$_2$ at room temperature. The mono-layer sample is different from the tri-layer and bi-layer samples. There are shade lines (crevasses) visible in the figure, and super-lattices are present in the yellow square marked D. These crevasses were not found in the bi-layer and tri-layer crystals. Fig. 5(b) shows the intensity profile along the red line in D. The ordered structures do not have long-range correlation and exist as domains. Figure 5(c) shows an enlarged view of the yellow frame in Fig. 5(a). The contrast of this image is reversed from Fig. 5(a) and thus the black points correspond to Ta atoms. The super-lattices shown by blue circles do not form a hexagonal structure (Fig. 5(c)). Figure 5(d) is the FTI of the yellow frame in Fig. 5(a). We confirm similar satellite patterns by performing FT at different location in Fig. 5(a). The properties of these satellites are shown in Table 1(c). The triple **q** vectors in the mono-layer break the threefold symmetry similar to the T-phase [14,15] (Fig. 5(e)).

3. **Discussion and Conclusion**

Here, we summarize the results presented in the previous sections. In the tri-layer sample, C-CDW clearly occurs even at room temperature. This result is surprising because the C-CDW phase does not appear in bulk samples above 220 K (Fig. 1(c)). In the bi-layer sample, satellites similar to the stripe phase in 2H-TaSe$_2$ are observed. Moreover, satellites similar to the T-phase in 1T-TaS$_2$ are present in the mono-layer sample. However, these phases are not observed at room temperature in bulk samples. In particular, the last two are never observed in the bulk samples without the cooling-heating cycle (Fig. 1(c)). To explain these results, we consider the following mechanisms.

First, the CDW in the tri-layer sample is C-type. This can be explained by the negative pressure effect. The CDW transition temperature ($T_{CDW}$) of MX$_2$ shows an increasing tendency with increasing interlayer distance [26]. In fact, the $T_{CDW}$ is increased by mechanical exfoliation in TiSe$_2$ [27]. The C-phase was not observed in reference [16-18] probably because their samples were not standing-free and the negative pressure was lost. However, evidences of the C-phase were reported in mono-layer at low temperature (not standing-free) [19] and in the surface of a thick 1T-TaS$_2$ sample[20] where the negative pressure effect takes a dominant role.

Second, we consider the bi-layer and mono-layer results. It is difficult to apply the above discussion to bi-layers and mono-layers because these phases should not exist under the conditions of this measurement (Fig. 1(c)). Previous studies have considered that stabilization of structures with anisotropic charge configurations such as the stripe phase (bi-layer) and the T-phase (mono-layer) can be explained by the inter-layer Coulomb interaction. The conventional Ginzburg-Landau (GL) equations [23, 24] support this idea. However, the stabilization of a structure such as the T-phase in mono-layer cannot be explained by the inter-layer Coulomb interaction. Therefore, we require a new mechanism.

Rotational-symmetry-breaking point defects such as vortices or disclinations were not discovered during a detailed survey of Fig. 5(a). There is a fundamental difference between defects in CDW and defects in superconductivity. In superconducting systems, a point defect such as an Abrikosov vortex attempts to be formed. On the other hand, in CDW systems, an in-plane line defect such as a domain wall attempts to be formed. In fact, domains are inevitably generated in the T-phase and the stripe phase. There is a difference between two domains in the CDW phase $\theta$ of the CDW MWF $\Psi = |\Psi| \exp(i\theta)$. Therefore, the entropy increases because the degree of freedom of the phase $\theta$ becomes large. In other words, a CDW does not need to generate vortices because it has topological defects equivalent to vortices in the KT phase from the start. Accordingly, we can conclude that there are no vortices or vortex-pairs in 1T-TaS$_2$.

To explain the structures in the bi- and mono-layer samples, it is necessary to reconsider triple **q**. This condition exists for the three-dimensional order of CDW [28]. If triple **q** is broken, the discrepancy between conventional studies and the bi- and mono-layer results is resolved, because traditional theories are based on triple **q** and three-dimensional crystals (bulk). An examination of Fig. 3(f), Fig. 4(e) and Fig. 5(e) reveals that triple **q** is broken with decreasing dimensionality. Moreover, the stripe structure in the mono-layer emerges because triple **q** is broken (Fig. 5(f)).

In conclusion, CDWs occur in tri-, bi- and mono-layer crystals. In a mono-layer sample, the CDW does not exhibit the KT phase that accompanies disclination type defects. Instead, we found CDW with domain wall type defect structures. This structure is a new triclinic stripe state without **q$_1$** + **q$_2$** + **q$_3$** = **0**. It is not necessary to maintain triple **q** in pure two-dimensional CDW systems (Table 1(c), Table 1(d)). Consequently, new states are created in the mono-layer and bi-layer samples. The stripe structure formed by breaking the triple **q** condition in the mono-layer may be useful for understanding other stripe structures such as copper oxides [29] and iron-based superconductors [30] in terms of anisotropic charge order. Moreover, a strong electron-phonon interaction forms shadow crevasses (Fig. 5(a), Fig. 5(f)). This shows that the electron-phonon interaction is of central importance in thinned samples. This idea is applied to systems in which an electron-electron interaction plays the most important role such as charge order in organic conductors [31]. New equations and models are needed if we are to realize pure two-dimensional systems. In addition, it is suggested that breaking of three-dimensional order causes new structures to be formed.

## 4. Methods
### 4.1 Sample Preparation

The single crystals of 1T-TaS$_2$ were grown in excess sulphur by the usual iodine vapor transport method. The prereacted powder of 1T-TaS$_2$ and a certain amount of excess sulphur were put in one end of a quartz tube and the tube was sealed in vacuum. The ampule was heat-treated in such a way that the mixture at one end of the quartz tube was at 950 to 830°C and the temperature of the other end was 70-80°C lower. It was found that single crystals were grown not only in the lower temperature end but also in the hotter one. The quartz tube was rapidly quenched into water to insure the retention of the 1T-phase.

### 4.2 Exfoliation

The thin samples are made using the exfoliation method. That is, the presence of the van der Waals gap with bonding makes it possible to exfoliate films of MX$_2$ with various thicknesses from its bulk in a similar manner to graphite. Some of the flakes were randomly chosen and cleaved using Scotch tapes and then transferred to transmission electron microscope microgrids following the method developed by Meyer and co-workers[32].

### 4.3 Scanning Transmission Electron Microscope

We used low voltage scanning transmission electron microscope (STEM). This method does not require a substrate and causes little or no damage to the samples[33]. Therefore, STEM is well suited for the measurement of ultrathin samples. We used this technique to observe formation of different CDWs in tri-layer, bi-layer, and mono-layer samples.

**Acknowledgements**

We thank Toshiro Tani, late Shoji Tanaka and late Takashi Sambongi for the instruction of sample preparation techniques and donation of some of the samples used in this experiment. We also thank the aforementioned professors, Koichi Ichimura, Toru Matruura and Junya Ishioka for stimulating discussions.

**Competing interests**

The authors declare no conflict of interest.

**Contributions**

K. S. and S. T. contributed to the original idea and supervised the project. S. T. and D. S. contributed sample preparation. Z. L., K. S., S. T. performed the experiment. D. S., Z. L. and S. T. analyzed the data. All authors contributed to the interpretation of the results. D. S., K. N. and S. T. wrote the manuscript and supplementary materials. All authors read and approved the final manuscript.

**Fundings**

Z. L. and K. S. acknowledge support from Grant-in-Aid for Scientific Research on Innovative Areas (MEXT KAKENHI Grant No. 25107003) and the JST Research Acceleration Program. This work was supported by JSPS KAKENHI (No. 26287069).

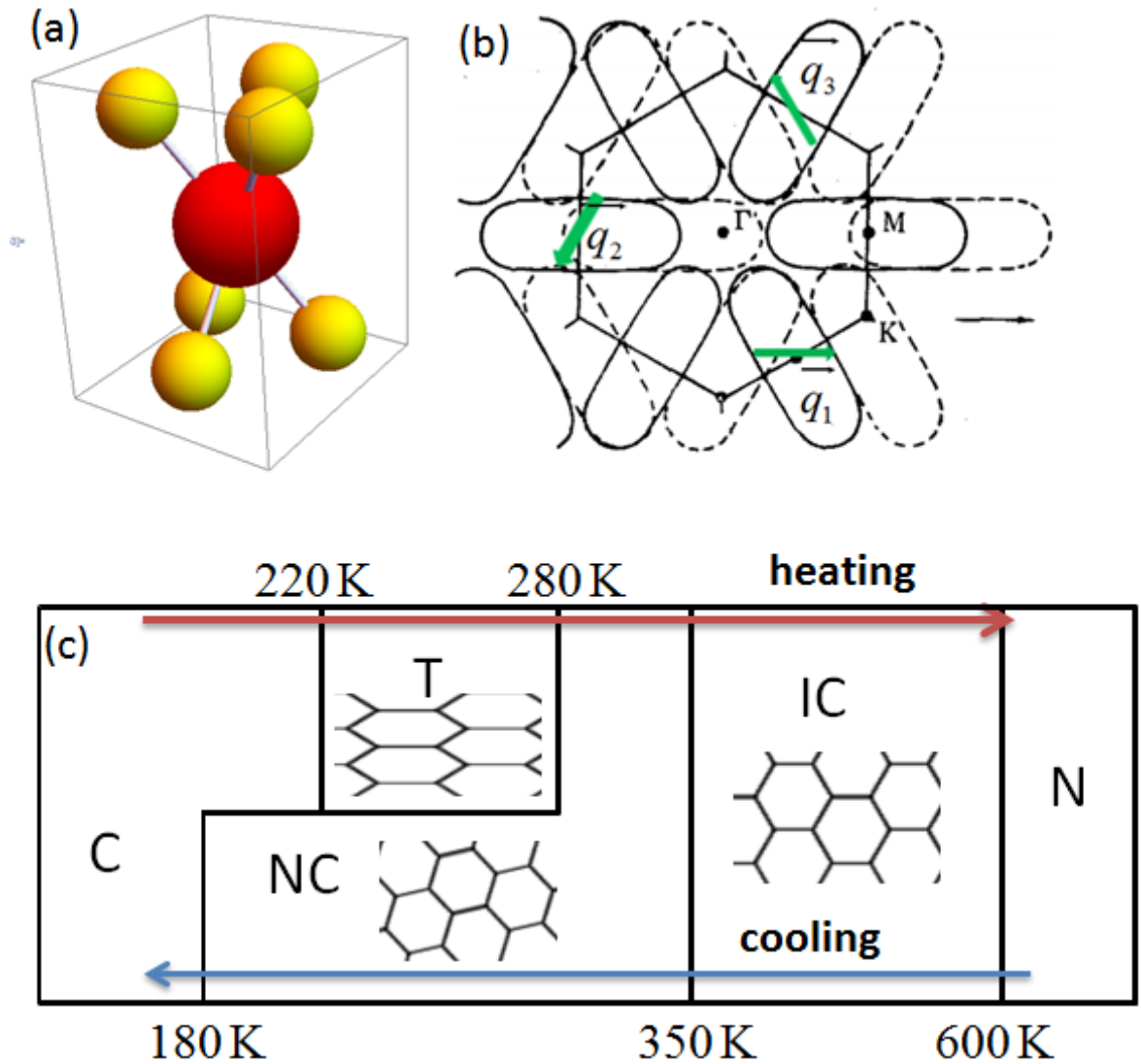

Figure 1: Properties of 1T-TaS$_2$
(a): 1T-polytype structure. The red spheres are tantalum atoms. The yellow spheres are sulfur atoms. (b): The Fermi surfaces of 1T-TaS$_2$ [12]. The nesting vector (green vectors) are mutually equivalent. The sum of the **q** vectors is **0** (Triple **q**). (c): CDW states depend on temperature in bulk 1T-TaS$_2$. The blue arrow represents the cooling cycle. The red arrow represents the heating cycle. C: Commensurate state [12], T: Triclinic state [14, 15, 21] which has stretched honeycomb lattices. Between 220 and 280K the T-satellites are observable but not the NC-satellite or the C-super-lattice reflection, even after thermal cycling. NC: Nearly commensurate state [12], IC: Incommensurate state [12], N: Normal metal state.

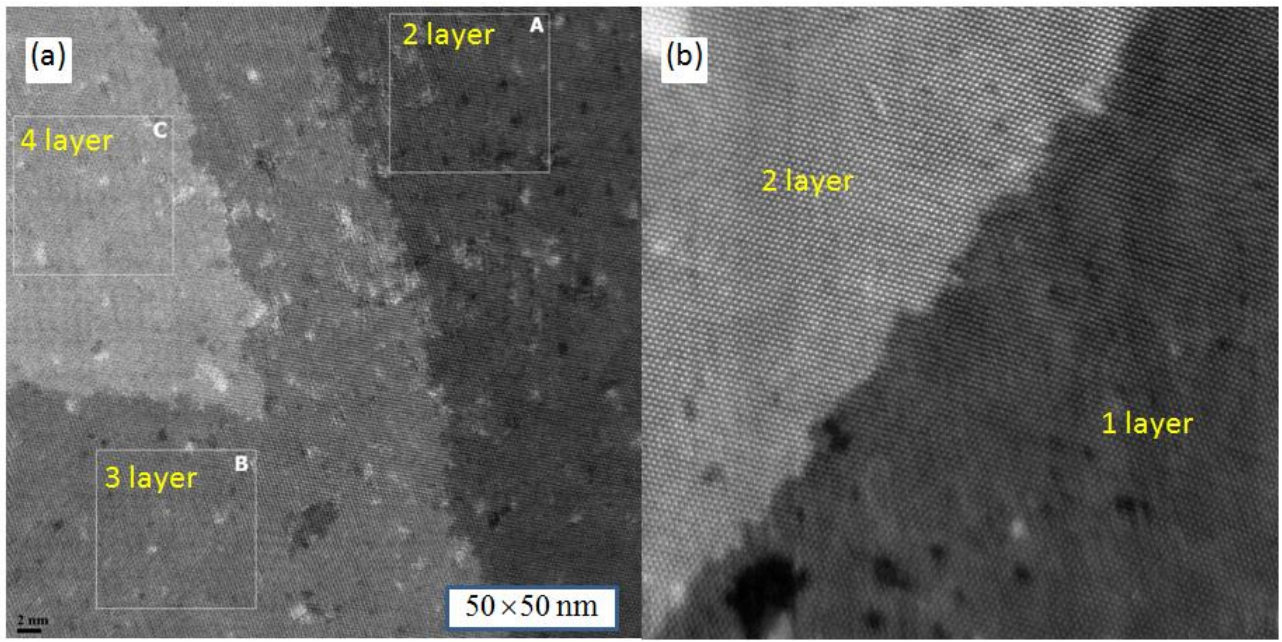

Figure 2: STEM image of a thin layer sample of 1T-TaS$_2$ including several layers. STEM captures images based on information obtained from scattered electrons, and thus a domain with a lot of scatterers is brighter. Thus, the magnitude of the brightness corresponds to the number of layers. (a) The darkest domain (right side of figure) is a bi-layer. The brightest domain (upper left side of figure) has four layers. In addition, the intermediate domain is a tri-layer. (b) The brightest domain has two layers, the darkest domain has one layer.

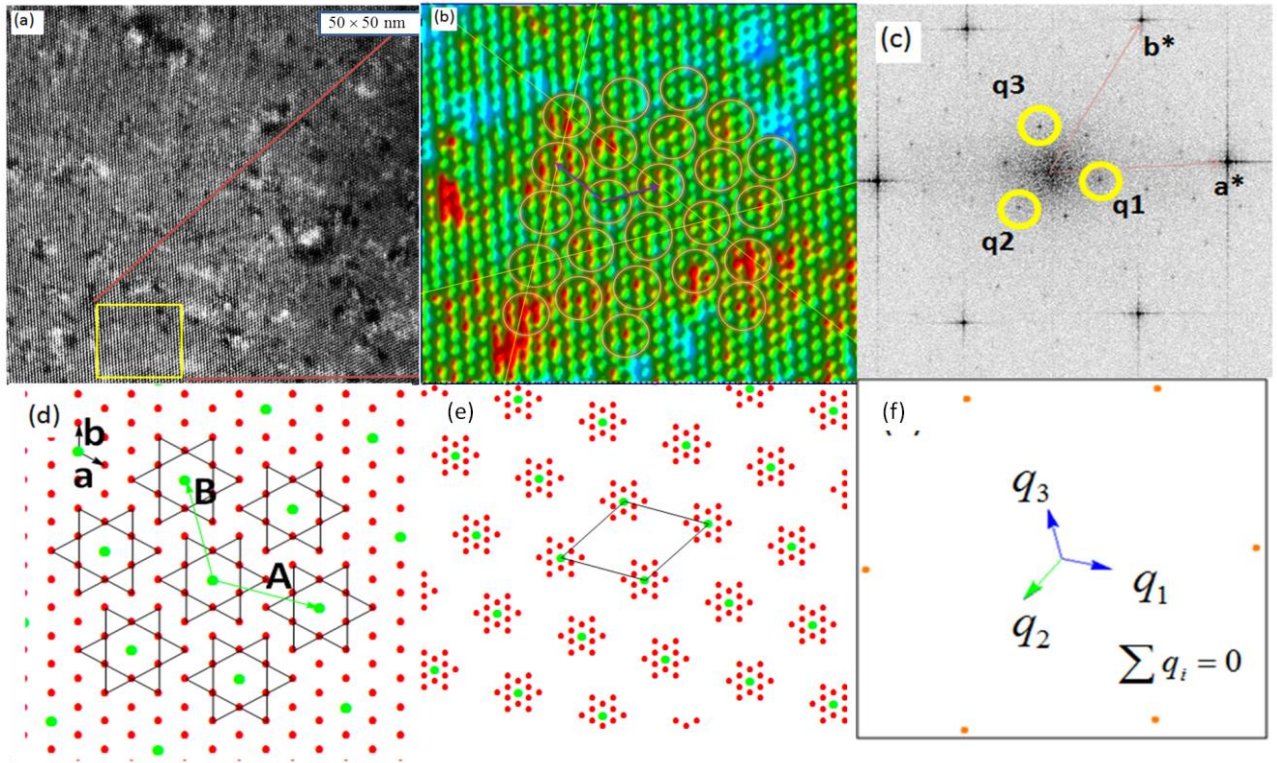

Figure 3: Tri-layer 1T-TaS$_2$

(a): STEM image of tri-layer 1T-TaS$_2$ at room temperature. Surprisingly, super-lattices of C-CDW are seen throughout the entire sample and are particularly obvious in the yellow frame in the lower left of this image. The super-lattices form hexagonal structures as seen in Fig. 3(e). (b): Top- view of the three-dimensional intensity plot of the area in the yellow frame in Fig. 3(a). Higher scattering magnitudes are shown in red. Brown circles show super-lattices. Purple vectors are super-lattice vectors (**A**, **B**). (c): Fourier transformed image (FTI) of Fig. 3(a). **q**$_1$, **q**$_2$, **q**$_3$ are C-CDW satellites. Satellite peaks are clearly present. This is due to an overlap of different diffraction orders. (d): A sketch of the tantalum layer of 1T-TaS$_2$ [25]. **a**, **b** are basic lattice vectors. The red points are Ta atoms and the green points are Ta atoms forming the center of the star-like super-lattice. **A**, **B** are super-lattice vectors in C-CDW (**A** = 4**a** + **b**, **B** = -**a** + 3**b**, |**A**| = |**B**| = √13|**a**|). (e): A sketch of super-lattices in C-CDW state [25]. The super-lattices form a new hexagonal lattice with basic vectors **A** and **B**. This is called a √13×√13 structure. (f) Satellite patterns of the tri-layer sample. The orange points are Bragg peaks of the lattice. The blue arrows are nesting vectors. The green arrow is -**q**$_1$-**q**$_2$, which overlaps with **q**$_3$, so the sum of **q** vectors is **0**.

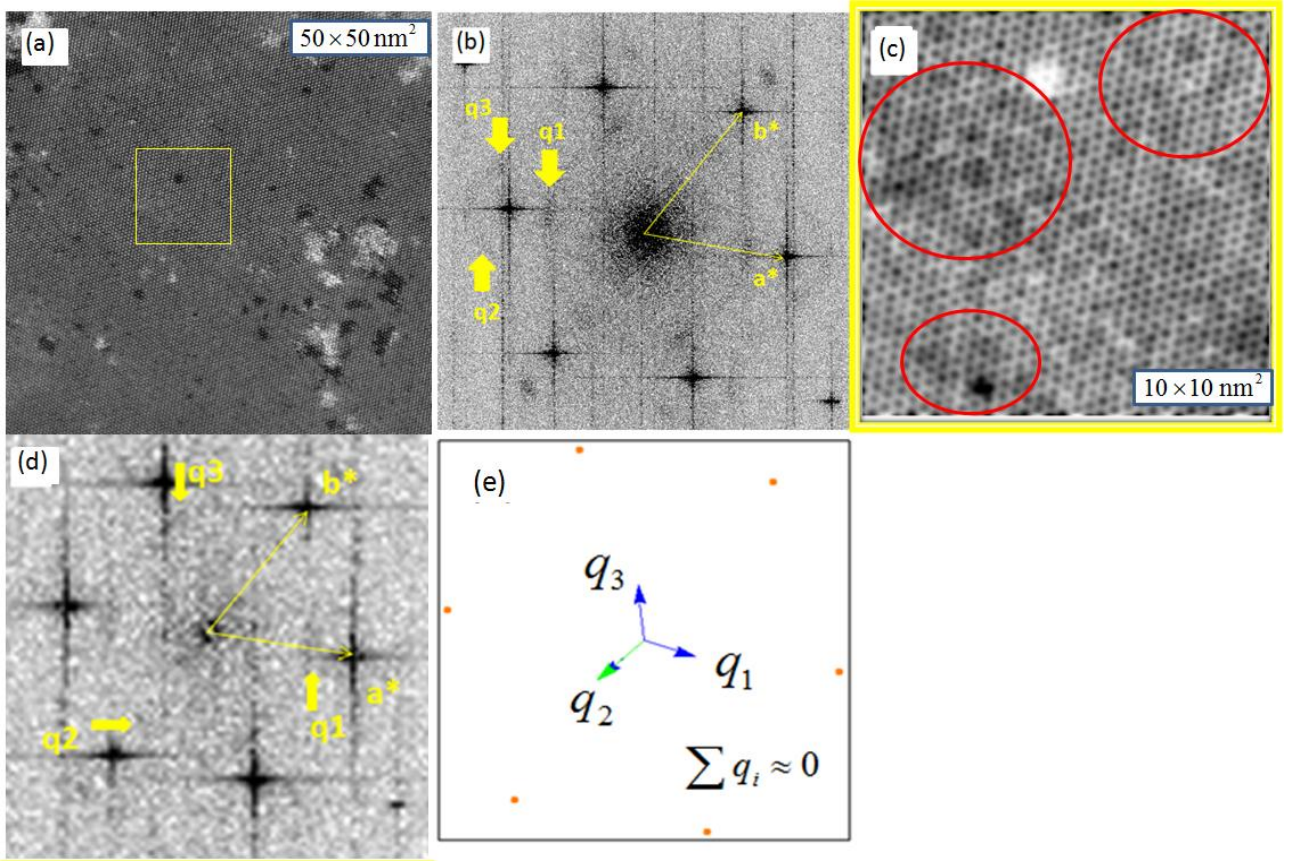

Figure 4: Bi-layer 1T-TaS$_2$
(a): STEM image of bi-layer 1T-TaS$_2$ at room temperature. Super-lattices of CDW are seen through- out the entire sample and are particularly obvious in the yellow frame in the middle of this figure. (b): FTI of Fig. 4(a). The CDW satellites shown by the yellow arrows are largely diffused. (c): Enlarged view of the yellow frame in Fig. 4(a) with inverted contrast. The brightest circle in Fig. 4(c) corresponds to the darkest circle in Fig. 4(a). Black hexagons correspond to super-lattices. Unlike in the tri-layer sample, super-lattices do not have clear hexagonal structures in the bi-layer. Local order structures, which are shown by red circles, are present in the bi-layer. (d): FTI of Fig.4(c). The CDW satellites shown by the yellow arrows are largely diffused. These are similar to those shown in Fig. 4(b). (f) Satellite patterns of the bi-layer sample. The orange points are Bragg peaks of the lattice. The blue arrows are nesting vectors. The breen arrow is -$q_1$-$q_2$ slightly deviates from $q_3$, so the sum of **q** vectors is approximately **0**.

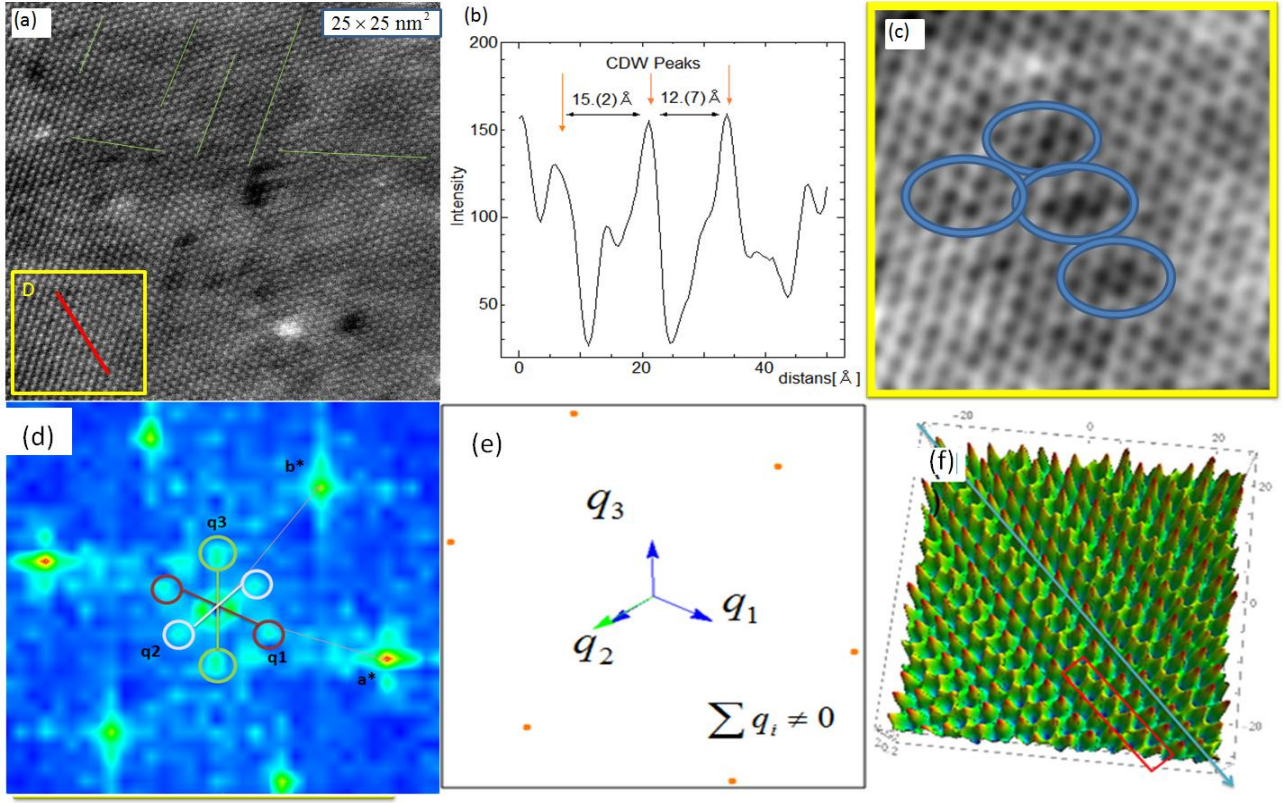

Figure 5: Mono-layer 1T-TaS$_2$
(a): STEM image of mono-layer 1T-TaS$_2$ at room temperature. There are shadow crevasses (Shown by green lines, see supplementary material 2 for an enlarged view with eyeguides). These correspond to areas with few scatterers. D shows domains with ordered structures. (b): An intensity plot along the red line in D. The orange arrows show super-lattices enhanced in Fig. 5(c). (c): An enlarged image of D in Fig. 5(a) with inverted contrast. The black hexagons shown by the blue ellipses are super-lattices. Super-lattices do not form honeycomb lattices. (d): FTI of D in Fig. 5(a). The transformed region is tiny, so the peaks are broad. However, we confirm the presence of triple **q** vectors. (e) Satellite patterns of the mono-layer sample. The orange points are Bragg peaks of the lattice. The blue arrows are nesting vectors. The green arrow is -**q**$_1$-**q**$_2$ which clearly differs from **q**$_3$, so triple **q** condition **q**$_1$+**q**$_2$+**q**$_3$=**0** is broken. (f) The charge configuration is reproduced from the satellites in Fig. 5 (e). The structures correspond to shadow crevasses (red frame) and stripe structure (blue arrow).

|  | Experimental Results | | Bulk Data (Previous Works) | |
|---|---|---|---|---|
| (a) | Tri-layer | | Commensurate phase | |
| **q** | $/\mathbf{q}_i\|/\|\mathbf{a}*\|$ | $\psi$ | $/\mathbf{q}_i\|/\|\mathbf{a}*\|$ | $\psi$ |
|  | 0.280(8) | 13.3° | 0.277 | 13.90 |
| (b) | Bi-layer | | Nearly-Commensurate phase | |
|  | $/\mathbf{q}_i\|/\|\mathbf{a}*\|$ | $\psi_i$ | $/\mathbf{q}_i\|/\|\mathbf{a}*\|$ | $\psi_i$ |
| $\mathbf{q_1}$ | 0.287(2) | 9.0° | | |
| $\mathbf{q_2}$ | 0.277(2) | 10.0° | 0.283 | 12.7° |
| $\mathbf{q_3}$ | 0.287(1) | 11.5° | | |
| (c) | Mono-layer | | T-phase ($T = 270K$) [15] | |
|  | $/\mathbf{q}_i\|/\|\mathbf{a}*\|$ | $\psi_i$ | $/\mathbf{q}_i\|/\|\mathbf{a}*\|$ | $\psi_i$ |
| $\mathbf{q_1}$ | 0.323(0) | 7.5° | 0.288 | 15.6° |
| $\mathbf{q_2}$ | 0.276(3) | 15.5° | 0.303 | 13.4° |
| $\mathbf{q_3}$ | 0.293(8) | 19.0° | 0.286 | 11.84° |

| (d) | $\mathbf{q_1 + q_2 + q_3}$ | Correlation Length |
|---|---|---|
| higher layer | = **0** | |
| tri-layer | = **0** | $70 \pm 25$ Å |
| bi-layer | ≈ **0** | $30 \pm 10$ Å |
| mono-layer | ≠ **0** | $7.5 \pm 2.5$ Å |

Table 1: ( a ) Parameters of **q** vectors for the Bi-layer sample: Two of the **q** vector lengths are the same. This is a feature of the stripe phase. The magnitude of the **q** vectors is shown due to the three-fold symmetry. (b) Parameters of **q** vectors for the mono-layer sample. Only the $\mathbf{q}_2$ value of the Nearly commensurate phase is shown due to the three-fold symmetry. (c) It is confirmed that the lattice symmetry is maintained. The symmetry of triple **q** is clearly broken. Correlation lengths were estimated by auto-correlation analysis